\documentclass[runningheads]{llncs}
\usepackage[T1]{fontenc}
\def\doi#1{\href{https://doi.org/\detokenize{#1}}{\url{https://doi.org/\detokenize{#1}}}}
\usepackage{graphicx}
\graphicspath{{./images/}}
\usepackage{hyperref}

\usepackage{listings}
\usepackage{xfrac} 
\usepackage{xcolor}
\usepackage{arydshln}
\usepackage{vcell}
\usepackage{float}	
\usepackage{cite}		
\usepackage{amsmath}	
\allowdisplaybreaks

\begin{document}
\title{Lightweight Encoder-Decoder Architecture for Foot Ulcer Segmentation}
\author{Shahzad Ali\inst{1}\orcidID{0000-0002-4949-8335} 
\and Arif Mahmood\inst{2}\orcidID{0000-0001-5986-9876} 
\and Soon Ki Jung\inst{1}\orcidID{0000-0003-0239-6785}}
\authorrunning{S. Ali et al.}
\institute{School of Computer Science and Engineering, Kyungpook National University (KNU), Daegu, South Korea\\ \email{shazadali@knu.ac.kr}, \email{skjung@knu.ac.kr}\\
\and Department of Computer Science, Information Technology University (ITU), Lahore, Pakistan\\ \email{arif.mahmood@itu.edu.pk}}
\maketitle

\begin{abstract}
Continuous monitoring of foot ulcer healing is needed to ensure the efficacy of a given treatment and to avoid any possibility of deterioration. Foot ulcer segmentation is an essential step in wound diagnosis. We developed a model that is similar in spirit to the well-established encoder-decoder and residual convolution neural networks. Our model includes a residual connection along with a channel and spatial attention integrated within each convolution block. A simple patch-based approach for model training, test time augmentations, and majority voting on the obtained predictions resulted in superior performance. Our model did not leverage any readily available backbone architecture, pre-training on a similar external dataset, or any of the transfer learning techniques. The total number of network parameters being around 5 million made it a significantly lightweight model as compared with the available state-of-the-art models used for the foot ulcer segmentation task. Our experiments presented results at the patch-level and image-level. Applied on publicly available Foot Ulcer Segmentation (FUSeg) Challenge dataset from MICCAI 2021, our model achieved state-of-the-art image-level performance of 88.22\% in terms of Dice similarity score and ranked second in the official challenge leaderboard. We also showed an extremely simple solution that could be compared against the more advanced architectures.

\end{abstract}

\section{Introduction} \label{SEC_intro}
Diabetes is a lifelong condition, and a diabetic person is at lifetime risk for developing foot ulcer wounds, which severely affects the life quality. Getting an infection further complicates the situation and may lead to limb amputations and even death. Such diabetic foot ulcer wounds need to be examined regularly, by the healthcare professionals, for diagnosis and prognosis, including assessing current condition, devising a treatment plan, and estimation of complete recovery accordingly.

\begin{figure}[h]
	\centering
	\includegraphics[width=\textwidth]{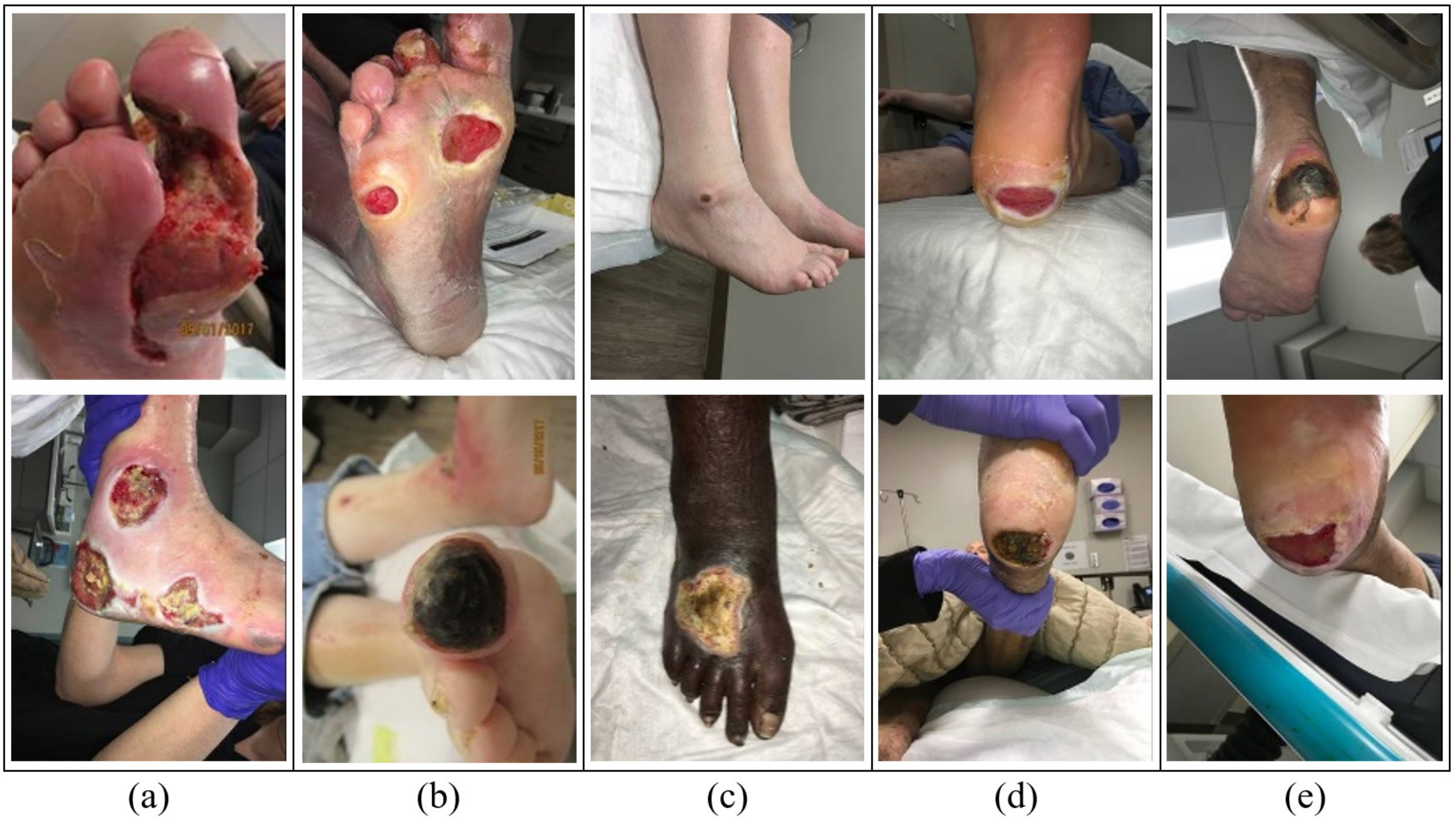}
	\caption{Typical challenging cases from the Foot Ulcer Segmentation (FUSeg) Challenge dataset: (a) heterogeneous wound shapes and their random positions, (b) color variations of wounds, (c) changes in skin tone, (d) background clutter, and (e) change in viewpoints. These images are cropped, and padding is stripped off for better display.}
	\label{FIG_1}
\end{figure}

Innovations in technology have resulted in better sensors and storage media thus, paving the way for advanced clinical procedures. The use of cameras and smartphones is getting common to obtain images of ulcer wounds each time a patient comes for an examination. The foot ulcer analysis is a lengthy process beginning from the visual inspection of wounds to determining their class type, severity, and growth over time by comparing past images side by side. Such subjective measures may cause human errors resulting, even with the utmost care, in an additional variability in enormously gathered data and hours of work in producing annotations. By utilizing artificial intelligence (AI) algorithms in general and deep learning (DL) techniques in particular, a vast amount of medical data is possible to process and analyze faster, accurately, and affordably. These algorithms are helping the healthcare industry to administer improved medical procedures, rapid healing, save huge expenses, and boost patient satisfaction. The segmentation is an essential step in a foot ulcer analysis pipeline. Having a reliable and efficient wound segmentation model could better aid in the evaluation of the condition, analysis, and deciding an optimal treatment procedure. 

The goal of foot ulcer wound segmentation is to label every pixel in an image either as wound \textit{(foreground)} or everything else \textit{(background}). There are several challenges in performing foot ulcer segmentation (as shown in Fig. \ref{FIG_1}) like heterogeneity in wound shape and color, skin color, different viewpoints, background clutter, lighting conditions, and capturing devices.

In this study, we propose an end-to-end lightweight deep neural network to perform foot ulcer wound segmentation which is robust to the challenges and generalizes well across the dataset without requiring any user interaction. Our model is inspired by the U-Net \cite{Ronneberger2015} and ResNet \cite{He2016} and includes the key features of both models. Each residual block in the proposed model has group convolution layers \cite{Krizhevsky2012} to keep the number of learnable parameters low. In addition, a residual connection, channel attention, and spatial attention are also integrated within each convolution block to highlight the relevant features and identify the most suitable channels to improve the prediction accuracy. The following are the main contributions of this study:

\begin{itemize}
	\item[--] We propose an end-to-end lightweight model for the foot ulcer wound segmentation primarily utilizing group convolutions.
	\item[--] Channel and spatial attention layers are combined with the residual connection within each block to form new \textit{residual attention} (ResAttn) block. There is no need to use standalone attention blocks resulting only in an increase in total trainable parameters and having a significant toll on overall model training time.	
	\item[--] We use test time augmentations (TTA) with the majority voting technique to get better segmentation results.
	\item[--] Experimental evaluation on publicly available Foot Ulcer Segmentation (FUSeg) dataset shows superior results. Our method stood second when compared with the top methods from the FUSeg Challenge leaderboard\footnote[1]{(\url{https://uwm-bigdata.github.io/wound-segmentation}) last accessed on Jan. 6, 2022.}.
\end{itemize}
The remainder of this paper is organized as follows. In Sect. \ref{SEC_related_work}, we provide an overview of the related work on the segmentation problem and attention techniques. Section \ref{SEC_proposed_method} describes our proposed model and experimental setup. Section \ref{SEC_experiments} presents the experimental details, results, and a brief discussion. Finally, the conclusion is given in Sect. \ref{SEC_conclusion}.

\section{Related Work} \label{SEC_related_work}
\subsection{Classical Segmentation Methods} 
Several probabilistic and image processing methods, machine learning, and deep learning techniques fall under this category. Edge detection, clustering, adaptive thresholding, K-means, and region-growing algorithms are a few well-known image processing methods used for segmentation \cite{Bahdanau2014}. These methods being not data hungry are fast, and most struggle to generate a reliable outcome for unseen data and thus fail to generalize their performance. Earlier machine learning algorithms typically made the best use of hand-crafted features based on image gradients, colors, or textures for segmentation. Such algorithms include classifiers such as multi-layer perceptron (MLP), decision trees, support vector machine (SVM) \cite{Wang2017}.

\subsection{Deep Learning-Based Segmentation Methods}
\textit{Convolution neural networks} (CNNs) have been successfully used for biomedical segmentation tasks such as segmenting tumors from breast, liver, and lungs using MRI and CT scans, nuclei segmentation in histological images \cite{Kumar2020,Caicedo2019}, skin lesion, polyp, and wound segmentation in RGB images \cite{Long2015,He2020,Wang2020}. Deep learning-based approaches have outperformed other approaches for foot ulcer segmentation \cite{Caicedo2019} since they are good to learn hidden patterns and generalize well for new data. Some well-known CNN-based architectures such as \textit{Fully Convolutional Neural Network} (FCN), \textit{U-Net}, \textit{Mask-RCNN}, and lightweight mobile architecture like \textit{EfficientNet} \cite{Ronneberger2015,Long2015,He2020} are utilized to perform wound segmentation in various studies \cite{Wang2020,Chino2020}.
 
\subsection{Attention Mechanisms}
These mechanisms allow a vision model to pay better attention to the salient features or regions in the input feature maps. This concept is closely related to image filtering in computer vision and computer graphics to reduce the noise and extract useful image structures. Bahdanau et al. \cite{Bahdanau2014} made the very first successful attempt to include the attention mechanism for an automated natural language translation task. 
\textit{Residual Attention Network} proposed by Wang et al. \cite{Wang2018} used non-local self-attention to capture long-range pixel relationships. Hu et al. \cite{Hu2018} used global average pooling operations to emphasize the most contributing channels in their proposed \textit{Squeeze-and-Excitation} (SE) blocks. Several other efforts have been made to incorporate spatial attention. Woo et. al \cite{Woo2018} made a notable effort with the \textit{Convolutional Block Attention Module} (CBAM). It consisted of the channel and spatial attention in a sequential fashion which led to significant improvement in the model representation power. Wu et al. \cite{Wu2021} proposed an \textit{Adaptive Dual Attention Module} (ADAM) that captured multi-scale features for recognizing skin lesion boundaries.

\section{Proposed Method} \label{SEC_proposed_method}
\subsection{Model Overview}
\begin{figure}[t]
	\centering
	\includegraphics[width=12cm]{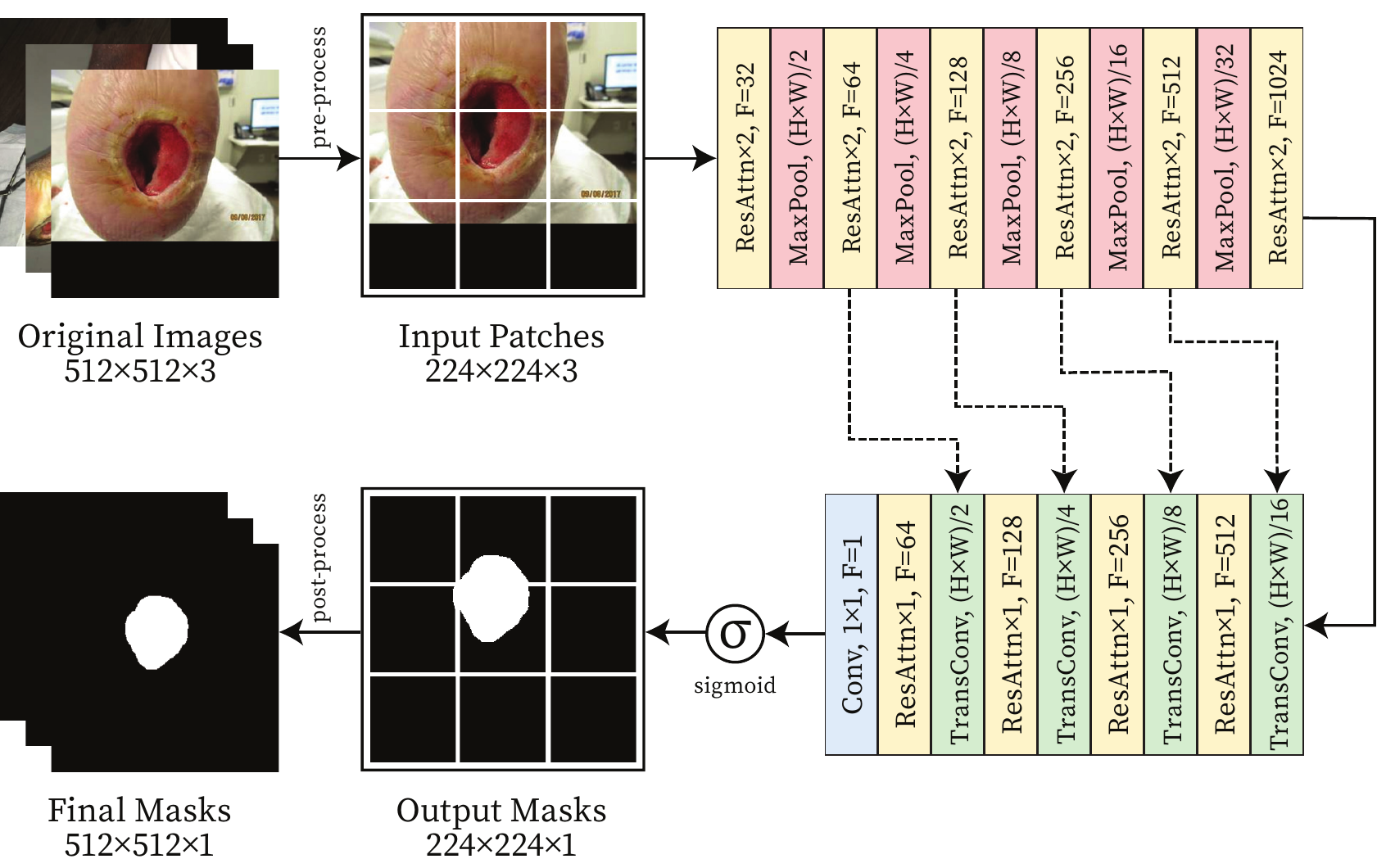}
	\caption{The proposed foot ulcer segmentation model is a U-shape model with redesigned convolution blocks as \textit{ResAttn} blocks. Final activation is sigmoid ($\sigma$), used instances of ResAttn are given in every block name (e.g., ``ResAttn×2'' means two block), $F$ indicates the number of output feature maps, and dotted lines represent a skip connection between encoding and decoding blocks. }
	\label{FIG_2}
\end{figure}

Our proposed model derives its key strength from the U-Net and ResNet architectures. We extended a U-shape model with the \textit{residual attention} (ResAttn) blocks. In each ResAttn block, convolution layers with variable receptive fields combined with channel and spatial attention better emphasize the contribution of meaningful features at different scales. Fig. \ref{FIG_2} shows the proposed architecture having two branches for image encoding and decoding purposes. Each branch contains a series of ResAttn blocks either with max-pooling or transpose convolution layers. Given an input image, feature extraction is performed during downsampling (encoding), followed by the reconstruction branch to upscale the feature maps (decoding) back to the input size. A series of transpose convolution layers upscales the element-wise summation of the feature maps. These feature maps come from the previous block and skip connections and thus have the same spatial size. The last ResAttn block outputs 64 channels that are reduced to 1 by a 1×1 convolution layer. Finally, a sigmoid function scales the dynamic range to [0,1] interval.

We carefully considered the impact of design choices. The \textit{point kernel convolutions} were preferred inside the ResAttn blocks since they require fewer training parameters than a convolution with a 3×3 or higher kernel. Our model initially produces 32 feature maps for each input RGB patch rather than 64 in the case of a standard U-Net. Since these feature maps grew twice in number by each encoding block, we saved a large amount of memory. Likewise, we found the \textit{group convolutions} extremely useful in remarkably reducing network parameters. The total trainable parameters of our model went down to 17\% as of its vanilla counterpart. We also observed that setting the value of \textit{groups} parameter to a multiple of 32 was sufficient for producing quality segmentation results.

\subsection{Loss Function}
In the training process, we used a linear combination of binary cross entropy loss $\mathcal{L}_{bce}$ and dice similarity loss $\mathcal{L}_{dice}$. The total segmentation loss $\mathcal{L}_{seg}$ was calculated as:
\begin{align}
	\mathcal{L}_{seg}&={\lambda_1\mathcal{L}}_{bce}+{\lambda_2\mathcal{L}}_{dice}, \label{EQ_1}\\
	\mathcal{L}_{dice}&=1-2\frac{\sum_{i}{g_ip_i}}{\sum_{i} g_i\sum_{i} p_i},\\
	\mathcal{L}_{bce}&=-\sum_{i}{(g_i\ln{\left(p_i\right)}+(1-g_i)\ln{(1-p_i)}),}
\end{align}
where $g$ is the ground truth binary mask, $p$ is the model prediction, $\lambda_1$ and $\lambda_2$ in Eq. \ref{EQ_1} are weighing parameters which were set to 1. The segmentation loss $\mathcal{L}_{seg}$ well trained our model and produced satisfactory segmentation performance. 

\begin{figure}[!b]
	\centering
	\includegraphics[width=9cm]{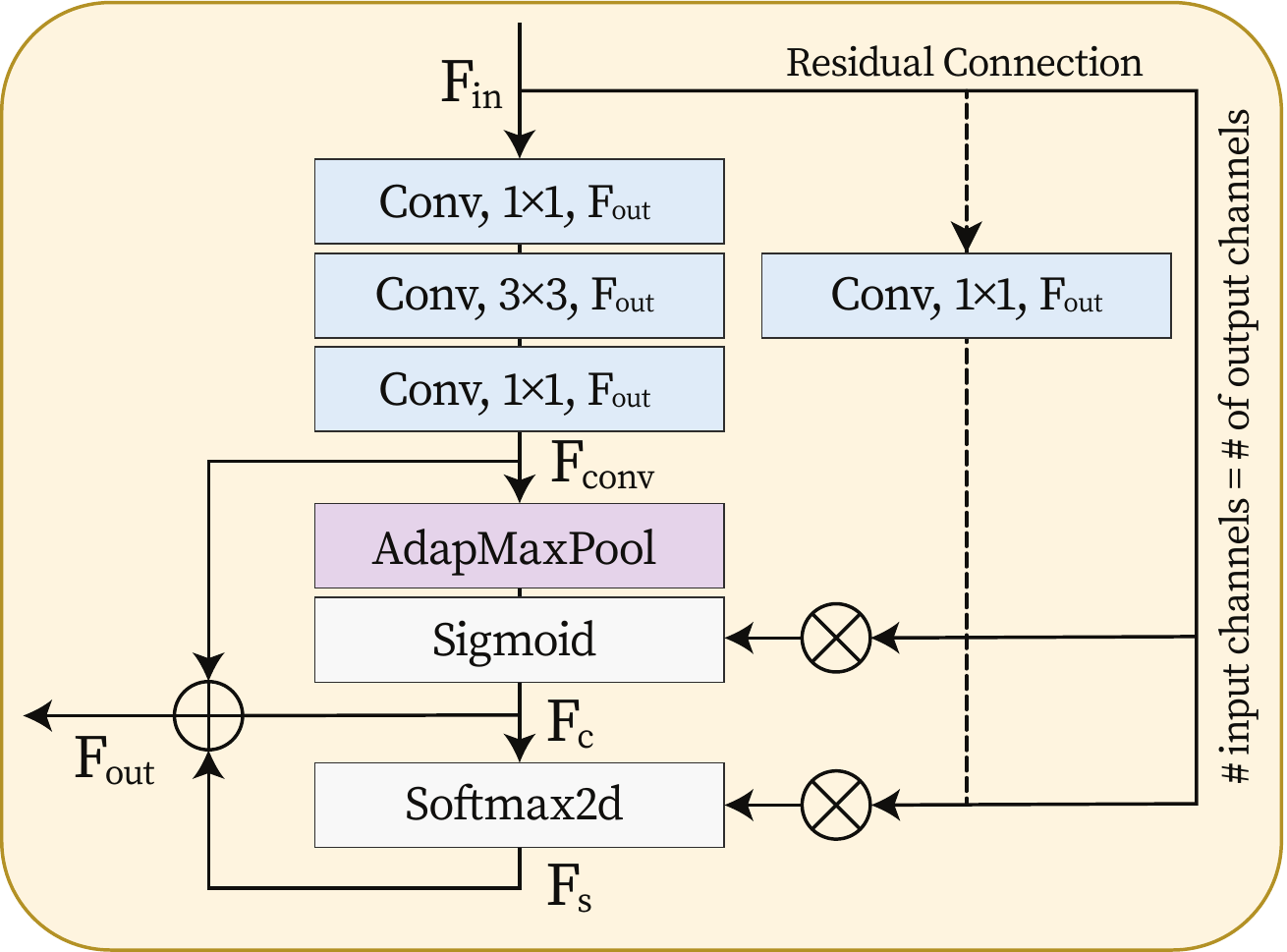}
	\caption{Residual attention (ResAttn) block in our lightweight model. Each convolution layer produces the same number of output feature maps ($F_{out}$). As long as the input and output channels are the same (i.e., $F_{in}=F_{out}$), the block input serves as a residual connection; otherwise, the dotted path is used. All convolution operations are followed by the batch norm and activation layers.}
	\label{FIG_3}
\end{figure}

\subsection{Residual Attention Block}
Each ResAttn block has three convolution layers with kernel sizes of 1×1, 3×3, and 1×1, respectively, along the main path. The fourth convolution layer with a 1×1 kernel serves as a \textit{residual connection} only when the number of input channels ($F_{in}$) is not equal to the number of output channels ($F_{out}$). All convolution layers are followed by the activation and batch norm layers. The total number of network parameters was reduced by choosing small and fixed-size kernels. Such small sized kernels reduce the effective receptive field resulting in the loss of spatial information. Furthermore, every pixel within a receptive field does not contribute equally to the output \cite{Luo2016}. This constraint can be alleviated by utilizing an attention mechanism to capture the global context information and improve the representation capability of extracted features. The spatio-channel attention as shown in Fig. \ref{FIG_3} remarkably increased the ability of model to pay attention to the meaningful task related information.

\begin{itemize}
	\item[--] \textbf{Channel Attention:} The channel attention vector $F_c\in\mathcal{R}^{(C\times1\times1)}$ 
	was obtained by squeezing the spatial dimension of an input feature map. We used adaptive max-pooling followed by a sigmoid function to get the probability estimate of the distinctiveness of each feature.
	\item[--] \textbf{Spatial Attention:} Unlike most spatial attention mechanisms proposed previously, we found that a 2D softmax over features to each spatial location was enough to yield a spatial map $F_s\in\mathcal{R}^{(1\times H\times W)}$. It attended the meaningful regions within the patches.	
\end{itemize}

Both attention maps were multiplied with the residual connection. It is either the block input or 1×1 convolution of the block input when the number of input channels were different from the number of output channels. Then their element wise summation with the output was obtained from the $conv-bn-gelu$ path. These operations can be expressed as Eq. \eqref{EQ_4} whereas the detailed scheme is given in Fig. \ref{FIG_3}.
\begin{align}
	\label{EQ_4}
	F_{out}=F_{conv}+\alpha F_c+\beta F_s,
\end{align}
where $F_{conv}$ is the output from the series of $conv-bn-gelu$, $F_c$ is the channel attention, $F_s$ is the spatial attention, and the two learnable weights are denoted as $\alpha$ and $\beta$. In our experiments, Gaussian Error Linear Units (GELU) were preferred over the Rectified Linear Unit (ReLU) for its stochastic regularization effect \cite{Hendrycks2016}. GELU activation function has shown promising results in the state-of-the-art architectures like GPT-3 \cite{Brown2020}, BERT \cite{Devlin2018}, and vision transformers \cite{Dosovitskiy2020}. 

\subsection{Experimental Setup}
We implemented our model in PyTorch \cite{Paszke2019} on a Windows 10 PC having an 8-core 3.6 GHz CPU and an NVIDIA TITAN Xp (12 GB) GPU. The training was carried out using input images cropped to 224×224 in a non-overlapped fashion. LAMB optimizer \cite{You2019} was used to update the network parameters with a learning rate of 0.001 and batch size of 16. The network was trained for 100 epochs only and, the epoch yielding the best dice similarity score was included in the results. No pre-training or transfer learning technique was used in any performed experiments except the Xavier weight initialization.

At test time, 224×224 sized patches of validation images were used to generate predictions. We used test time augmentations (TTA) \cite{Simonyan2015} at patch-level. Such augmentations included horizontal/vertical flips and random rotation by the multiple of ${90}^o$. We did not use multi crops at test time because the quality gain was negligible over the increase in computation time. The majority voting technique was used to decide the label at the pixel level.

\section{Experiments} \label{SEC_experiments}
\subsection{Dataset}
This dataset was released for the \textit{Foot Ulcer Segmentation Challenge} at the International Conference on Medical Image Computing and Computer Assisted Intervention (MICCAI) in 2021 \cite{Wang2022}. It is an extended version of the chronic wound dataset and has 810 training, 200 validation, and 200 test images. The size of images was kept fixed at 512×512 pixels by applying zero-padding either at the left side or bottom of the image. The ground truth masks for the test images were held private by the organizers for the final evaluation of challenge participants so we evaluated the model performance for validation images only. We employed online data augmentation transformations including horizontal/vertical flips, multiple random rotate by ${90}^o$, and random resized crops with high probability ($p\sim1.0$). Other augmentations of significantly low probability ($p\sim0.3$) included randomly setting HSV colors, random affine transformations, median blur, and Gaussian noise.

\subsection{Evaluation Metrics}\label{SUBSEC_evaluation}
The quality of predicted segmentation masks was evaluated comprehensively against the ground truth using five different measures such as Dice similarity index (DSC), Jaccard similarity index (JSI), sensitivity (SE), specificity (SP), precision (PR), which are defined as:
\begin{align}
	DSC&=\frac{2TP}{2TP+FP+FN},\\
	JSI&=\frac{TP}{TP+FP+FN},\\
	SE&=\frac{TP}{TP+FN},\\
	SP&=\frac{TN}{TN+FP}, \text{ and}\\
	PR&=\frac{TP}{TP+FP},
\end{align}
where TP, FN, TN, and FP represent the number of true positive, false negative, true negative, and false positive respectively. The output values of all these measures range from 0 to 1, and a high score is desired. Before evaluating the model performance, all obtained predictions were first binarized using a threshold value of 0.5.

\subsection{Comparison with Baseline Model}
We evaluated all model predictions obtained for the validation data on both patch-level and image-level for a fair comparison with other methods. A standard U-Net was trained from scratch, keeping the training configuration and augmentation transformations close to the original paper \cite{Ronneberger2015}, gave the best dice score of 89.74\% as compared to 91.18\% achieved by our lightweight architecture as shown in Table \ref{TAB_1}. The total number of parameters and the total number of floating-point operations per second (FLOPS) were significantly reduced to 16\% of the vanilla U-Net model. The first column in Table \ref{TAB_1} has the total network parameters in millions, the second column is for \textit{giga-floating-point operations per second} (GLOPS), and the rest of the columns present the performance metrics given in section \ref{SUBSEC_evaluation}.

\begin{table}[t]
	\centering
	\caption{Architecture and performance comparison (in terms of \%) between at the patch level. The best results shown in bold.}
	\label{TAB_1}
	\begin{tabular}{l|l|l|l|l|l|l|l} 
		\hline
		Model              & Param(M)$\downarrow$  & GFLOPS$\downarrow$\textit{ } & DSC$\uparrow$            & JSI$\uparrow$            & SE$\uparrow$             & SP$\uparrow$             & PR$\uparrow$              \\ 
		\hline
		U-Net
		(vanilla)  & 31.03         & 30.80               & 89.74          & 81.39          & 89.01          & \textbf{99.73} & \textbf{90.48}  \\ 
		\hline
		Proposed
		method & \textbf{5.17} & \textbf{4.9}        & \textbf{91.18} & \textbf{83.79} & \textbf{92.99} & 99.69          & 89.44           \\
		\hline
	\end{tabular}
\end{table}

\begin{figure}[!b]
	\centering
	\includegraphics[width=\textwidth]{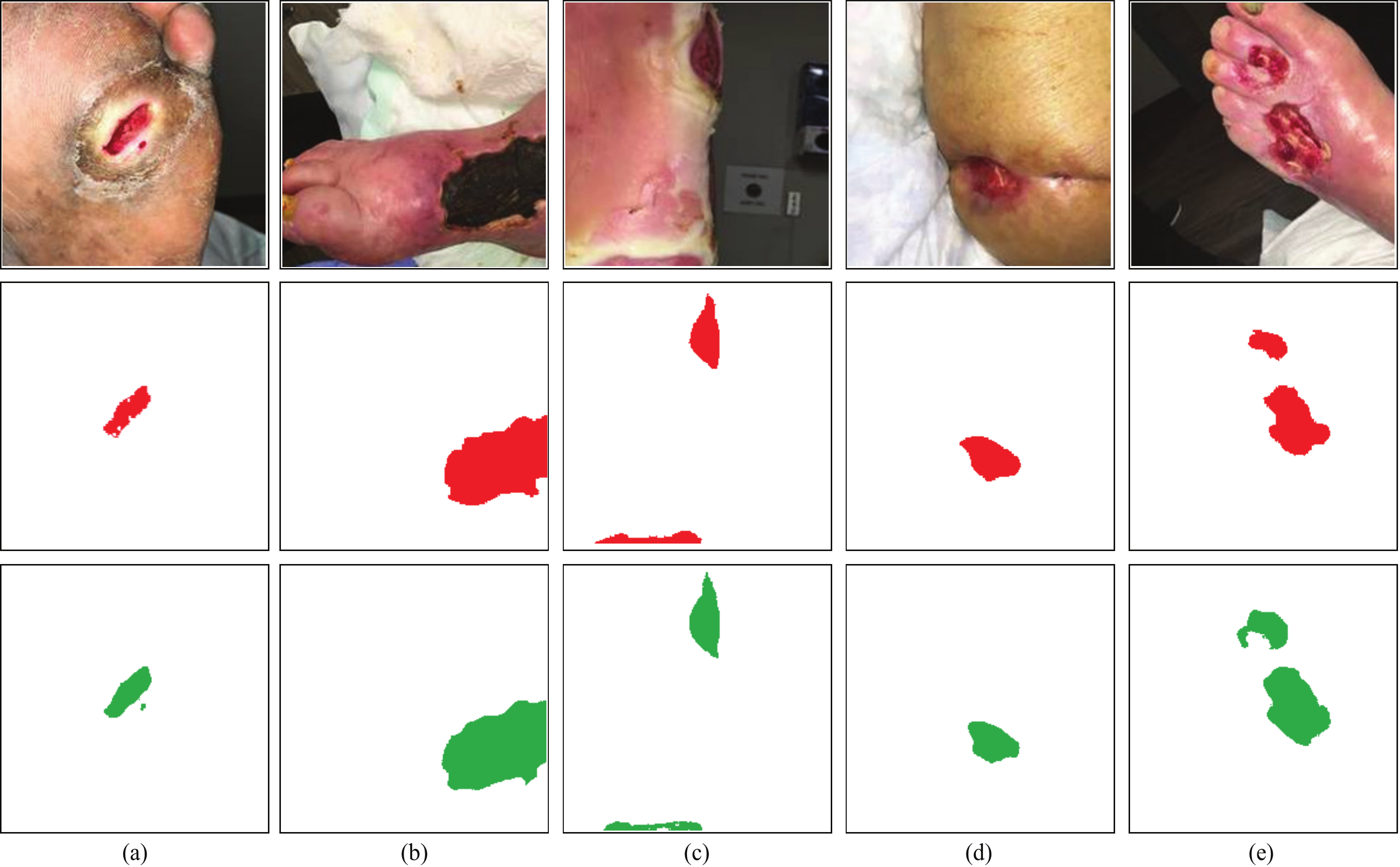}
	\caption{Example patches from the images of FUSeg validation data (\textit{top row}), ground truth masks in red color (\textit{middle row}), and segmentation prediction obtained from the proposed model in green color (\textit{last row}).}
	\label{FIG_4}
\end{figure}

Some example of patches extracted from the validation dataset images are shown in Fig. \ref{FIG_4}. The predicted segmentation results were almost identical to the ground truth masks. In some cases, as in Fig. \ref{FIG_4} (c) and (d), the model showed sensitivity to fresh wounds since they were high in color contrast in comparison to their surroundings. Fig. \ref{FIG_4} (a) represents a case where the model exhibited poor performance in capturing the fine-grained details potentially due to the extremely low number of learnable parameters.

\subsection{Comparison with Challenge Records}
For image-level evaluations, all 224×224 patch-level predictions were unfolded to recover the original image of size 512×512. The statistical results of our method for the validation images are given as Table \ref{TAB_2} in comparison with the participating teams in the challenge. Our method ranked second on the leaderboard and successfully competed with other wider and deeper architectures. These models often utilized pre-trained backbone in a U-shape architecture along with extensive ensemble approaches.

\begin{table}[!t]
	\centering
	\caption[]{The leaderboard of MICCAI 2021 Foot Ulcer Segmentation (FUSeg) Challenge. Our proposed method achieved the second-best place.}	
	\label{TAB_2}
	\begin{tabular}{c|l|l|c}
		\hline \# & Team & Model & DSC$\uparrow$ \\ \hline 
		1  & \begin{tabular}[c]{@{}l@{}}Amirreza Mahbod, Rupert Ecker, Isabella \\Ellinger~\textcolor{darkgray}{(Medical University of Vienna,}\\\textcolor{darkgray}{TissueGnostics GmbH)}\end{tabular}  & U-Net+LinkNet & 0.8880 \\ \hdashline[1pt/1pt] ~  & \textbf{Proposed method} & \begin{tabular}[c]{@{}l@{}}\textbf{U-Net with residual }\\\textbf{attention blocks}\end{tabular} & \textbf{0.8822} \\ \hdashline[1pt/1pt] 
		2  & \begin{tabular}[c]{@{}l@{}}Yichen Zhang~\textcolor{darkgray}{(Huazhong University of }\\\textcolor{darkgray}{Science and~Technology)}\end{tabular} & \begin{tabular}[c]{@{}l@{}}U-Net with HarDNet68 \\as encoder backbone\end{tabular} & 0.8757 \\ \hline 
		3  & \begin{tabular}[c]{@{}l@{}}Bruno Oliveira \\\textcolor{darkgray}{(University of Minho)}\end{tabular} & ~- & 0.8706 \\ \hline 
		4  & \begin{tabular}[c]{@{}l@{}}Adrian Galdran~\\\textcolor{darkgray}{(University of Bournemouth)}\end{tabular} & Stacked U-Nets~~~~~~ & 0.8691 \\ \hline 
		5  & \begin{tabular}[c]{@{}l@{}}Jianyuan Hong, Haili Ye, Feihong Huang,\\~Dahan Wang~\textcolor{darkgray}{(Xiamen University of }\\\textcolor{darkgray}{Technology)}\end{tabular} & ~- & 0.8627 \\ \hline
		6  & \begin{tabular}[c]{@{}l@{}}Abdul Qayyum, Moona Mazher, Abdesslam\\~Benzinou, Fabrice~Meriaudeau \textcolor{darkgray}{(University }\\\textcolor{darkgray}{of~Bourgogne Franche-Comté)}\end{tabular} & ~-  & 0.8229 \\ \hline
		7  & Hongtao Zhu \textcolor{darkgray}{(Shanghai University)} & U-Net with ASPP & 0.8213 \\ \hline
		8  & Hung Yeh \textcolor{darkgray}{(National United University)} & ~ & 0.8188 \\ \hline
	\end{tabular}
\end{table}

\section{Conclusion} \label{SEC_conclusion}
The use of deep learning methods for automated foot ulcer segmentation is the best solution to the laborious annotation task and analysis process. We proposed using ResAttn block based on the residual connection, spatial attention, and channel attention. Our lightweight architecture, with ResAttn blocks, outperformed several recent state-of-the-art architectures at the leaderboard of the Foot Ulcer Segmentation Challenge from MICCAI 2021. In addition, this study offers an alternative perspective by showing how minor yet highly valuable design choices can lead to excellent results when using a simple network architecture.

\subsubsection{Acknowledgment.} This study was supported by the BK21 FOUR project (AI-driven Convergence Software Education Research Program) funded by the Ministry of Education, School of Computer Science and Engineering, Kyungpook National University, Korea (4199990214394).

\bibliography{refs}
\bibliographystyle{splncs04}
\end{document}